\documentclass[prl,superscriptaddress,showpacs, twocolumn]{revtex4-1}

\usepackage{graphicx}
\usepackage{amsmath}
\usepackage{latexsym}
\usepackage{comment}
\usepackage{bm}
\usepackage{gensymb}
\usepackage{hyperref}
\usepackage{color}
\hypersetup{colorlinks,urlcolor=blue}
\usepackage[font=small,labelfont=bf,
   justification=centerlast ]{caption} 
\usepackage{subcaption}
\usepackage{csquotes}
\usepackage{amssymb}
\usepackage{tikz}
\usetikzlibrary{shapes,snakes}
\usepackage[section]{placeins}
\usepackage{bibentry}

\newcommand{\avg}[1]{\ensuremath{\left\langle #1 \right\rangle}}

\renewcommand{\vec}[1]{\bm{#1}}

\graphicspath{{./Figures/}}

\begin{document}
\nobibliography*

\title{Theory of rotational columnar structures of soft spheres}
\date{\today}
\author{J. Winkelmann} 
\affiliation{School of Physics, Trinity College Dublin, The University of Dublin, Ireland}
\author{A. Mughal}
\affiliation{Institute of Mathematics, Physics and Computer Science, Aberystwyth University, Penglais, Aberystwyth, Ceredigion, Wales, SY23, United Kingdom}
\author{D.B. Williams} 
\affiliation{School of Physics, Trinity College Dublin, The University of Dublin, Ireland}
\author{D. Weaire} 
\affiliation{School of Physics, Trinity College Dublin, The University of Dublin, Ireland}
\author{S. Hutzler} 
\affiliation{School of Physics, Trinity College Dublin, The University of Dublin, Ireland}

\begin{abstract}
There is a growing interest in cylindrical structures of hard and soft particles.
A promising new method to assemble such structures has recently been introduced by Lee \textit{et al.} [T.~Lee, K.~Gizynski, and B.~Grzybowski, Adv. Mater. \textbf{29}, 1704274 (2017)]. They used rapid rotation around a central axis to drive spheres of lower density than the surrounding fluid towards this axis. This resulted in different structures as the number of spheres is varied. 
Here we present comprehensive \textit{analytic} energy calculations for such self-assembled structures, based on a generic soft sphere model, from which we obtain a phase diagram.
It displays interesting features, including peritectoid points.
These analytic calculations are complemented by preliminary numerical simulations for finite sample sizes with soft spheres.
A similar analytic approach could be used to study packings of spheres inside cylinders of fixed dimensions, but with a variation in the number of spheres. 
\end{abstract}

\maketitle


Columnar crystals appear both in nature and in man-made objects and include biological systems such as viruses, flagella or microtubules \cite{charbonneau2018, erickson1973tubular, brinkley1997microtubules, amir2012dislocation} and nanotubes \cite{chopra1995boron, smalley2003carbon, sanwaria2014helical}.
A more remote but stimulating context for this research is the discovery of ``tubular crystals'' in simulations; these consist of various columnar structures interwoven in three dimensions \cite{douglass2017stabilization}.
 
Bubbles, emulsion droplets, hydrogel or plastic spheres are common constituents of such structures in laboratory experiments.
They can crystallise spontaneously when confined in cylindrical tubes \cite{lohr2010helical, meagher2015experimental, yamchi2015helical}.
For hard spheres the densest arrangement depends critically on the ratio of $D$, the cylinder diameter, to $d$, the sphere diameter, as was found in computer simulations \cite{mughal2011phyllotactic, mughal2012dense, mughal2013screw, mughal2014theory, fu2016hard, fu:2016wm}.
Such simulations have also been carried out for the packings of soft spheres \cite{wood2013self, winkelmann2017simulation, hysteresis2018}.

Recently, the subject has been given a new twist by the development of a novel experimental method, in which the rapid spinning of a liquid-filled column containing spheres of lower density than the surrounding fluid drives these toward the central axis \cite{lee2017non}.
Most of the data presented by Lee \textit{et al.} \cite{lee2017non} are for effectively \textit{hard} spheres.
This results in the familiar sequence of columnar hard sphere packings, but without the intervening ``line slips'', by which such structures are accommodated in response to the change in diameter $D$ of a densely packed cylinder \cite{winkelmann2017simulation}.
Numerical simulations have been performed to reproduce these observations, with some success \cite{lee2017non}.

\begin{figure}[h!]
\begin{center}
\includegraphics[width=1.0\columnwidth]{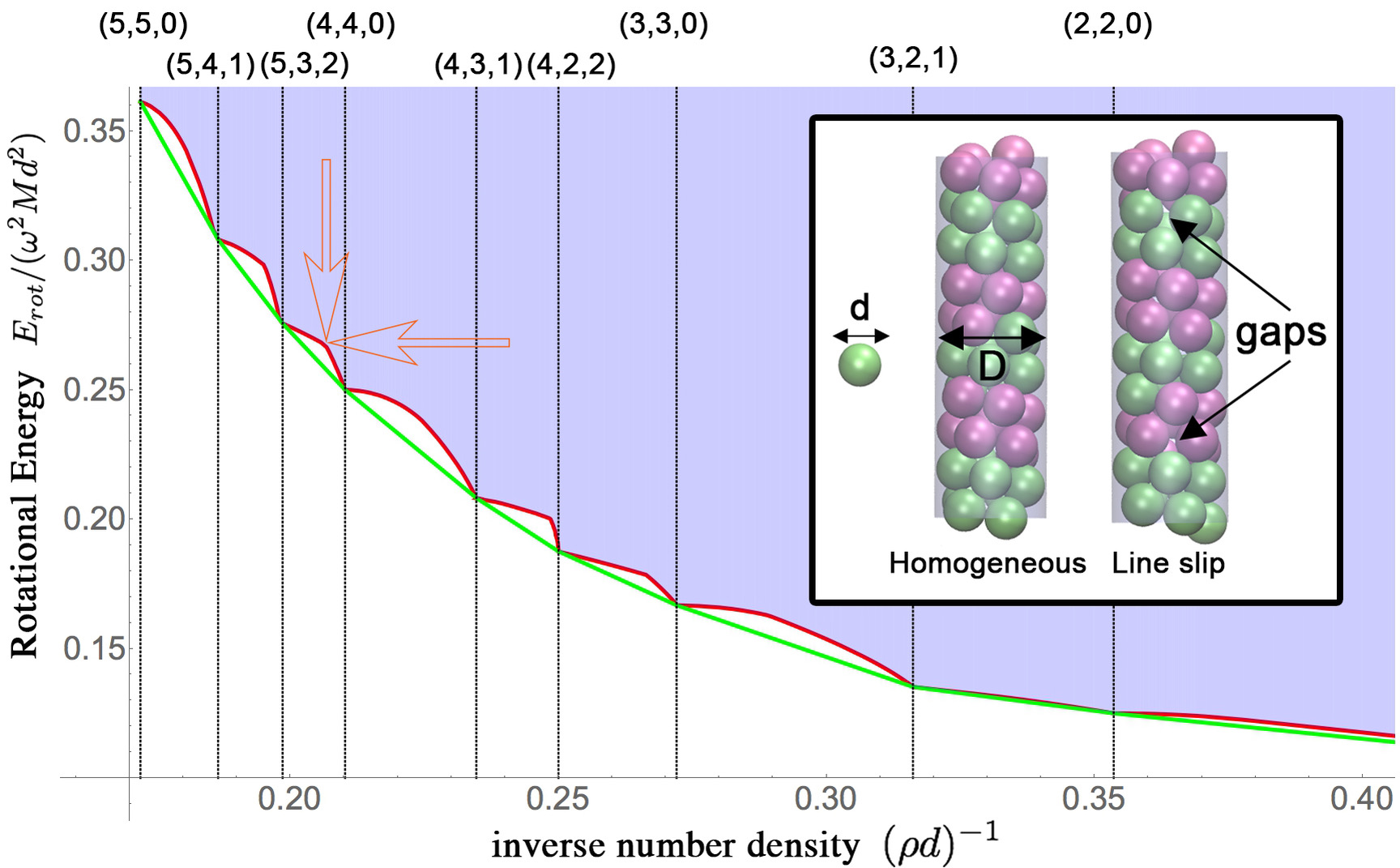}
\caption{
Minimal rotational energy $E_{\text{rot}}$ as a function of dimensionless inverse number density $(\rho d)^{-1}$ for \textit{hard sphere} packings.
The energy of the line-slip packings is given by the solid red line.
Vertical black lines indicate the location of homogeneous structures, identified by indices $(l,m,n)$.
The \textit{straight} solid lines (green line) between adjacent homogeneous structures indicates that binary mixtures have a lower energy compared to line-slip packings.
The shaded area is the region of all possible hard sphere packings which have a single value of distance $R$ from the central axis, by definition this precludes binary mixtures.
The inset shows examples of a homogeneous and a line-slip structure.
See text for an interpretation of the two red arrows in the main figure.
}
\label{fig2}
\end{center}
\end{figure}

Here we adopt a different approach using analytical methods to obtain a comprehensive phase diagram for such rotating columnar structures.
In practical terms, the two axes of such a phase diagram correspond to a variation of the linear number density $\rho$ (number of spheres per unit length), which can be varied in an experiment, and the squared rotational frequency $\omega^2$.
These are appropriately rescaled to be represented by dimensionless quantities, as indicated below.

We first investigate the hard sphere limit (corresponding to $\omega \rightarrow 0$ in an experiment).
We then perform analytic energy calculations for a system of soft spheres, where the total energy of the system is the sum of the rotational energy and an overlap energy (representing the elastic energy stored in the system due to contacts between adjacent spheres).
Finally, we compare our analytic results with computer simulations of soft spheres, where finite sample size introduces modifications to the analytical phase diagram.

\begin{figure}[t]
\begin{center}
\includegraphics[width=\linewidth]{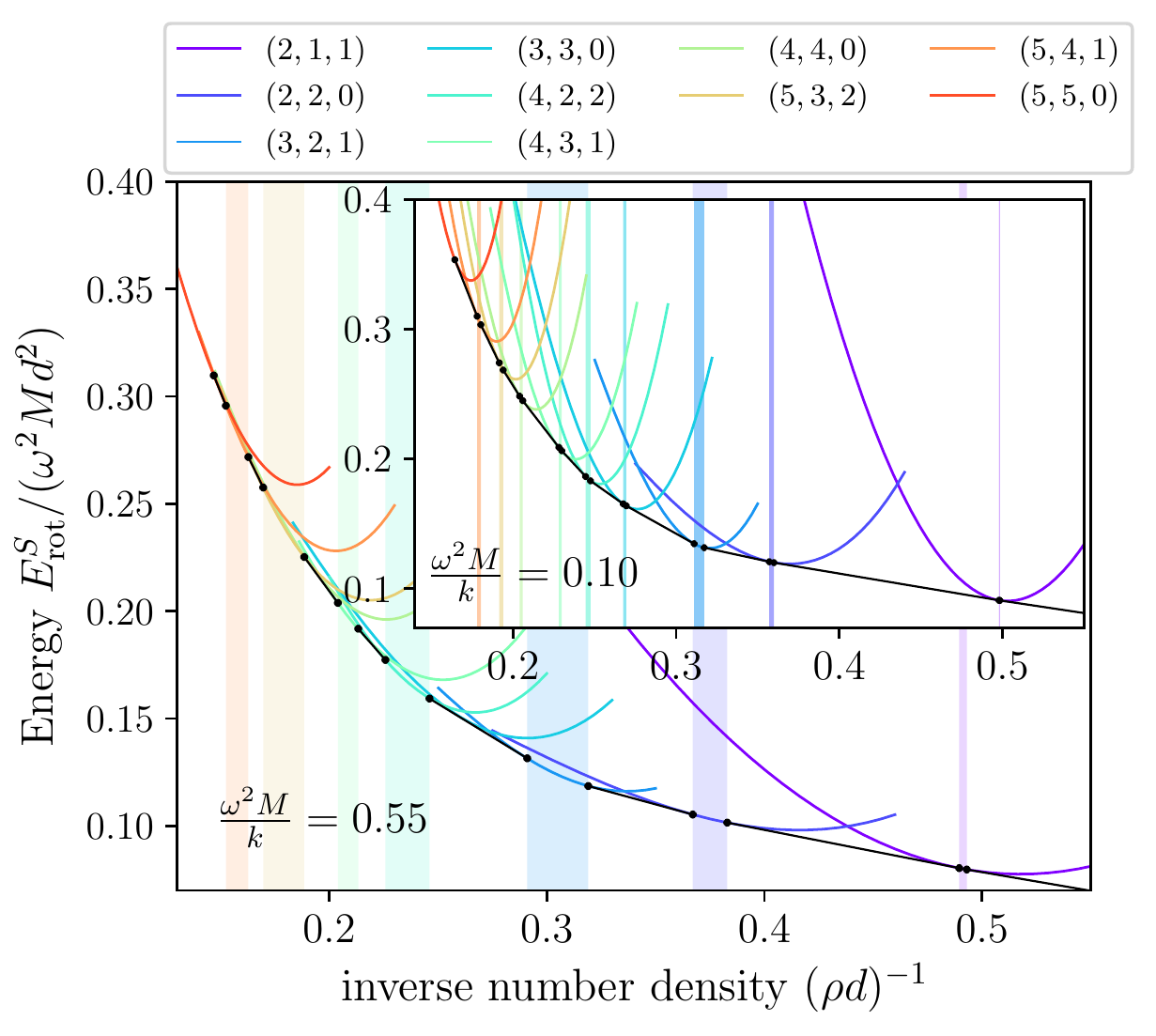}
\caption{
Minimal energy per sphere for all homogeneous structures as a function of dimensionless inverse number density $(\rho d)^{-1}$ for the case of \textit{soft spheres}, with harmonic interactions (see text).
The main figure shows the result for $\omega^2 M / k = 0.55$. The inset shows the energies for $\omega^2 M / k = 0.10$, which is close to the hard sphere limit. Common tangents between adjacent homogeneous structures are shown by the black lines and tangent points are shown by black dots. The range over which homogeneous structures are expected are highlighted by shaded strips (coloured according to the appropriate homogeneous structure - see key). Outside these ranges the common tangents have a lower energy and binary mixtures are expected.
}
\label{fig:energy}
\end{center}
\end{figure}

\emph{Hard spheres:}
Mughal \textit{et al.}  \cite{mughal2011phyllotactic, mughal2012dense, mughal2013screw, mughal2014theory} have computed the maximal-density columnar structures of hard spheres inside a cylinder.
While there is no mathematical proof of these results, they have been corroborated by others \cite{fu:2016wm, fu2016hard} and are in little doubt.
The packing fraction $\Phi$ of the densest structures was computed as a function of the ratio $D/d$ of cylinder to sphere diameters (Fig. 3 of Ref. \cite{mughal2012dense}).
Up to $D/d\approx 2.7$ these structures include only spheres in contact with the confining cylinder, so that all spheres are of the same distance $R$ from the central axis.
All of the homogeneous structures considered here are of this type.

The densest hard sphere packings can be classified as either {\it homogeneous} (previously called symmetric structure of single phase \cite{mughal2012dense, winkelmann2017simulation, hysteresis2018}) packings or {\it line-slip} arrangements; examples of both are shown in the inset of Fig.~\ref{fig2}.
For a homogeneous arrangement each sphere is in the same relation to six neighbouring spheres.
Such structures can be classified using the {\it phyllotactic} notation, i.e. a triplet of positive integers $(l=m+n,m,n)$ with $m\geq n$.
The point pattern formed by the sphere-cylinder contact features three families of spirals;
$l$, $m$, and $n$ count the number of spirals in each direction until the spirals repeat  \cite{mughal2014theory}.

Intervening between these homogeneous packings, which are found at particular values of $D / d$, are line-slip structures in which contacts are lost along a line separating two spiral chains of the homogeneous structure \cite{winkelmann2017simulation}.

We can transform known hard sphere packing results to give us the lowest (rotational) energy per sphere (that is, the minimal $R$) as a function of the dimensionless inverse number density $(\rho d)^{-1}$ (see Fig \ref{fig2}).
If the rotational velocity is $\omega$, then by the parallel axis theorem the rotational energy per sphere is $E_{\text{rot}}= \frac{1}{2} \omega^2 (I_0 + M R^2)$, where $I_0 = Md^2 / 10$ is the moment of inertia of a sphere with mass $M$.
Since the moment of inertia of a sphere is independent of $R$, we will omit this term in the calculations below.
For hard sphere packings the sphere centres are located at a distance $R =(D-d)/2$ from the cylindrical axis, thus the rotational energy for hard sphere packings is given by $E^{H}_{\text{rot}} = \frac{1} {8} \omega^2 M (D - d)^2$.
The dimensionless inverse number density $(\rho d)^{-1}$ can be computed from the diameter ratio and the packing fraction $\Phi$ as $(\rho d)^{-1} = \frac{2}{3} (\frac{d}{D})^2 \Phi^{-1}$.

Figure \ref{fig2} implies that the homogeneous structures, which occur for special values of $(\rho d)^{-1}$, minimise the rotational energy per sphere and we expect to observe a single-phase structure at these values of $(\rho d)^{-1}$.
In the intervening ranges, however, binary mixtures (consisting of two-phase structures) of the adjacent homogeneous structures are expected because the energies of these structures, dictated by the usual Maxwell (common tangent) construction, lies below the line slip energies in Fig \ref{fig2}.

Note that the hard sphere limit may be approached in two ways, as indicated by the red arrows in Fig. \ref{fig2} -- horizontally, as in the packing simulations of Ref. \cite{winkelmann2017simulation, hysteresis2018}, or vertically, as in the present Rapid Communication.

The resultant structure sequence is in accord with the findings of Lee \textit{et al.} \cite{lee2017non}.
In the following we address its modification in the case of soft spheres and finite columns.

\emph{Soft spheres. Analytical results.} The effect of moving away from the hard sphere limit is to widen the range in which the homogeneous $(l,m,n)$ structures are found.
This can be quantified and understood in terms of a transparent analytical description, illustrated by Fig \ref{fig:energy} and described below.

For a rotating column of soft spheres the energy per sphere is given by $E_{\text{rot}}^S=E_{rot}+E_{o}$, where the second term is due to their interaction.
The interaction energy associated with two contacting spheres $i$ and $j$ is given by $E_{ij}=\frac{1}{2}k\delta^2_{ij}$.
It is zero if their centres are more than $d$ apart.
Here $k$ is the spring constant and the overlap $\delta_{ij}=|\vec{r}_i-\vec{r}_j|-d$ depends on the sphere positions $\vec{r}_i$ and $\vec{r}_j$.
The overlap energy per sphere $E_{o}=\frac{1}{2}k\avg{\delta^2_{ij}}$ is obtained by summing over all pairwise interactions and dividing by the total number of spheres in the structure.
Thus the energy per sphere is given by,
\begin{equation}
\frac{E_{\text{rot}}^S}{M \omega^2 d^2} = \frac{1}{2} \frac{R^2}{d^2} + \frac{1}{2} \frac{k}{M \omega^2} \avg{\left(\frac{\delta_{ij}}{d}\right)^2}\,.
\label{eq:totenergy}
\nonumber
\end{equation}

Homogeneous structures are comprised of packings for which each sphere is in an identical relationship to every other sphere in the packing. Each sphere is at a distance $R$ from the central axis and is in contact with six neighbouring spheres.
From these constraints it follows that for a given number density $\rho$ the energy of a homogeneous structure can be varied \emph{only} by a uniform radial compression/expansion or twist, if it is to remain homogeneous (as we assume here).
Thus the energy per sphere for all homogeneous structures is an analytic expression to be minimised with respect to only two variables: $R$ and a twist angle $\alpha$ \cite{mughal2012dense}.
In the case of achiral structures this twist angle is zero due to symmetry, and only one free variable remains.

The minimised energies of all homogeneous structures as a function of the inverse number density $(\rho d)^{-1}$ are shown for two different values of $\omega^2 M / k$ in Fig \ref{fig:energy}.
These calculations are based on the full harmonic interactions between neighbours (see below for justification).
The common tangents between adjacent curves are shown by the black lines and the point of contact between the tangent and the curve by black dots. Where the common tangents are below the energy curves of the homogeneous structures, the binary mixtures are more stable. For the other values of $(\rho d)^{-1}$, highlighted by the shaded strips, a homogeneous phase is predicted. Low values of $\omega^2 M / k$ correspond to the hard sphere limit (see the inset of Fig \ref{fig:energy}); with increasing $\omega^2 M / k$ the overlap between spheres increases, resulting in a broadening of the range over which homogeneous structures are observed.

There is no loss of contacts in the range in which homogeneous structures are predicted.
This justifies the simplification that resulted from not taking loss of contacts into account, i.e. using the full harmonic approximation of interactions.

\begin{figure}[h!]
\begin{center}
\includegraphics[width=\columnwidth]{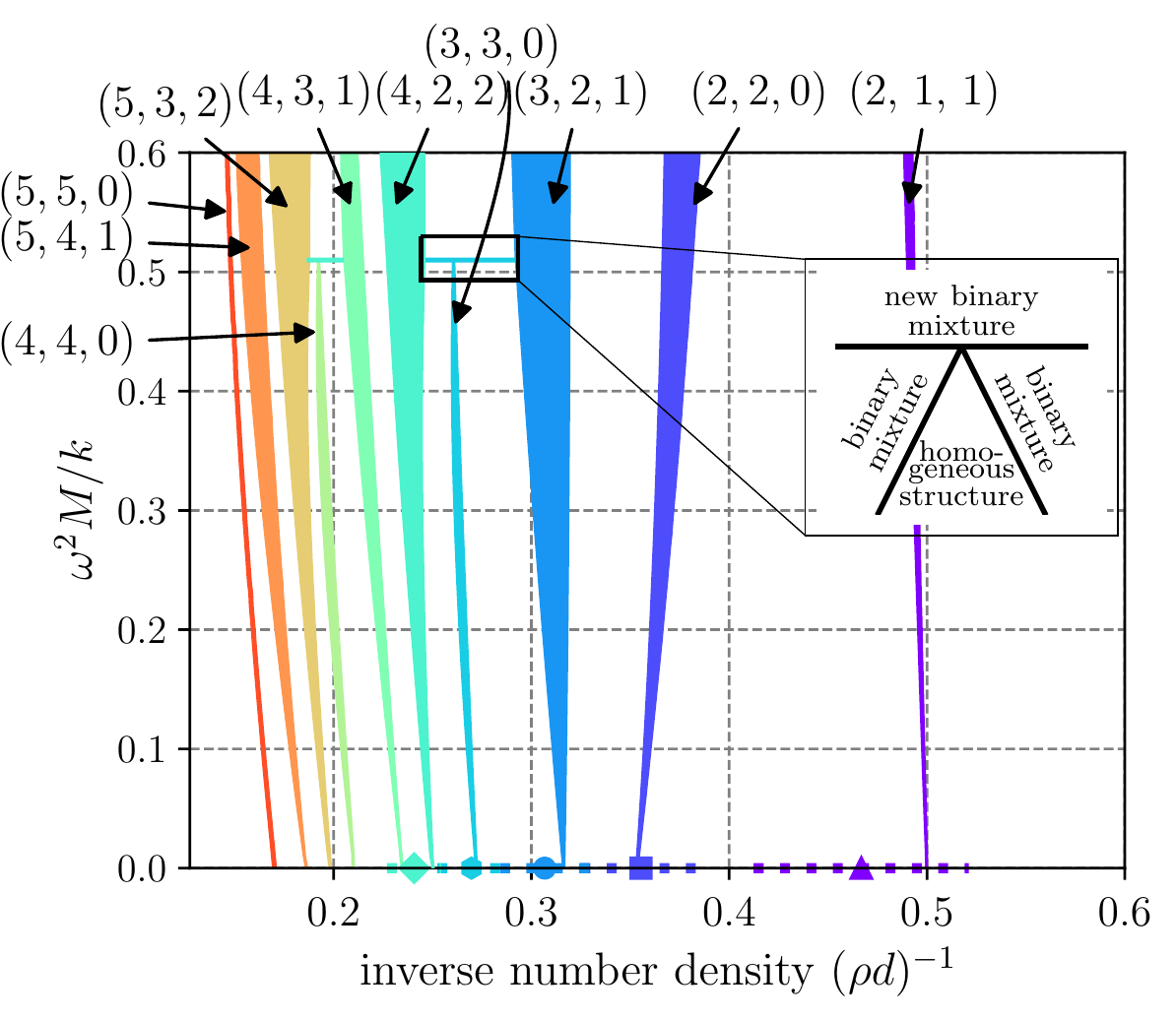}
\caption{
Calculated phase diagram of homogeneous structures (coloured, labelled regions) and their binary mixtures of adjacent homogeneous structures (intervening white space).
Most homogeneous regions expand with increasing $\omega^2 M / k$ (i.e. rotation frequency) but the achiral $(3, 3, 0)$ and $(4, 4, 0)$ structures vanish in \textit{peritectoid} points (see the inset).
In the case of the $(5,5,0)$ phase only the right hand boundary can be shown, since it is the last structure without inner spheres.
The $(\rho d)^{-1}$ values for the homogeneous structures are in good agreement with simulation results from Ref. \cite{lee2017non}, indicated by points with dashed horizontal error bars in the hard sphere limit.
}
\label{fig:phasediagram}
\end{center}
\end{figure}

The corresponding analytically calculated phase diagram is shown in Fig \ref{fig:phasediagram}.
In the hard sphere limit ($\omega^2 M / k \rightarrow 0$), the values of $(\rho d)^{-1}$ for the homogeneous structures are consistent with those from the simulations of Lee \textit{et al.} [Fig 3(b) in \cite{lee2017non}] which we have indicated by points with dashed horizontal error bars.
Lee's experimental data are shifted toward higher values of $(\rho d)^{-1}$.
This is possibly due to vibrations, keeping the hard spheres slightly apart, which gives them an effective diameter larger than their actual size.

With increasing $\omega^2 M / k$ the ranges of homogeneous structures expand, as expected. The upper part of the phase diagram is, however, much richer in detail than anticipated.
Of special interest is the vanishing of the homogeneous achiral structures $(3, 3, 0)$ and $(4, 4, 0)$ at a rotational velocity of $\omega^2 M / k \approx 0.5$. For high $\omega^2 M / k$ the achiral structures cannot compete with chiral structures: The latter can deform by twisting, while the former cannot.  

At the values of $\omega^2 M / k$ where these achiral structures disappear, there are \textit{peritectoid} points (see inset of Fig \ref{fig:phasediagram}).
The homogeneous structures vanish in a point and also their adjacent binary mixtures disappear.
The phase boundaries of the adjacent homogeneous structures show a change in slope where the new mixed structures appear.
This is due to the change of the common tangent, now to be taken between the second-nearest homogeneous structures.

\begin{figure}[t]
\begin{center}
\includegraphics[width=0.97\columnwidth ]{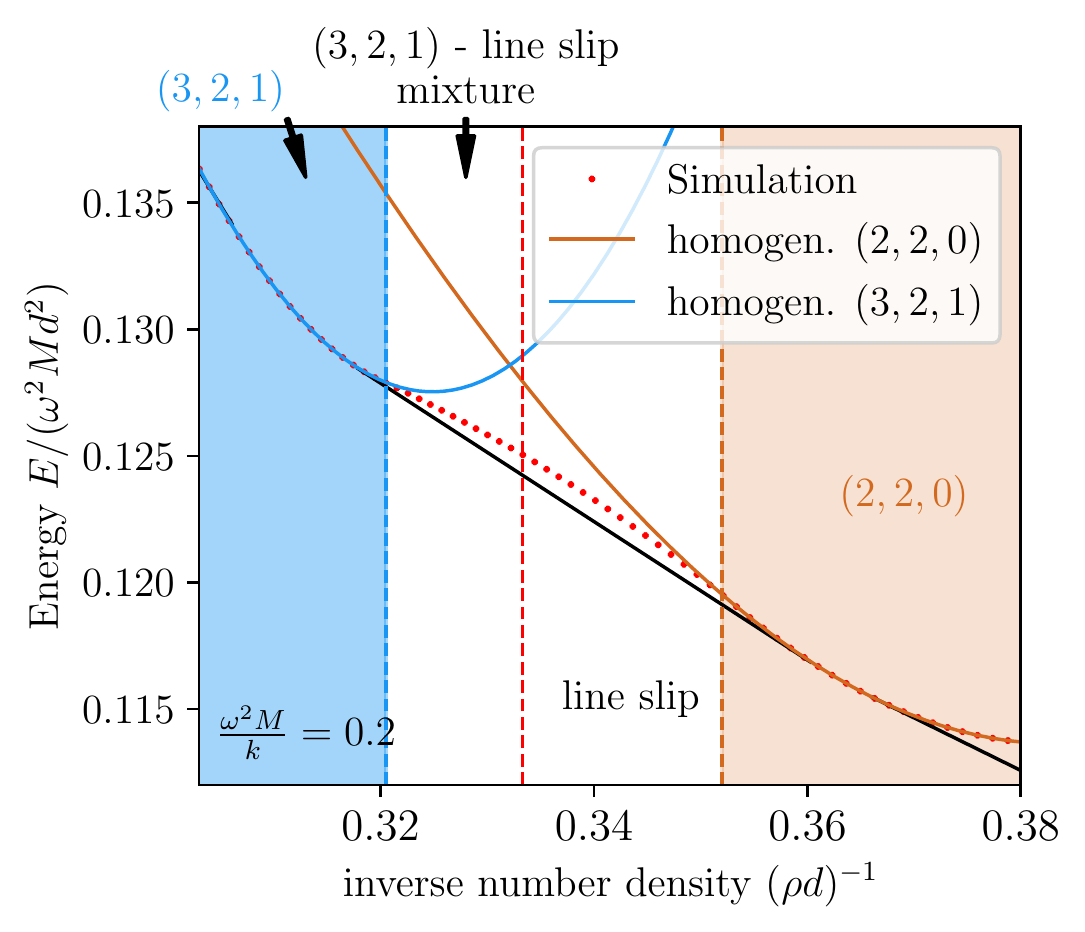}
\caption{
The red dots show the energy per sphere from finite-size simulations, for $\omega^2 M / k = 0.2$ and $N = 24$ spheres. 
Brown and blue solid lines show the previously analytically calculated energies for the homogeneous $(2, 2, 0)$ and $(3, 2, 1)$ structures, and the common tangents (black solid lines) represents the binary mixture of the adjacent structures.
At the vertical dashed blue line the homogeneous structure transforms into a binary mixture, whose energy is higher than that of the common tangent (due to finite-size effects).
Within the vertical dashed red and brown line line-slip structures are observed.
}
\label{fig:EnergiesSim}
\end{center}
\end{figure}

\emph{Numerical optimisation:}
The soft sphere model can be used in more general numerical simulations for a finite system of $N$ spheres which can occupy any position in a simulation cell of length $L$.
In doing so, we have again applied twisted boundary conditions \cite{mughal2012dense}.
Values of $N$ which are multiples of $12$ are compatible with  the homogeneous phases $(2, 1, 1)$, $(2, 2, 0)$ and $(3, 2, 1)$ and their corresponding line slips: 
in the simulations presented here we have used $N = 24$.
We use the Basin-Hopping method \cite{basinhopping} to search for the structure of minimum energy for particular values of number density $\rho = N / L$ and $\omega^2 M / k$.

\begin{figure}[t]
\begin{center}
\includegraphics[width=0.97\columnwidth ]{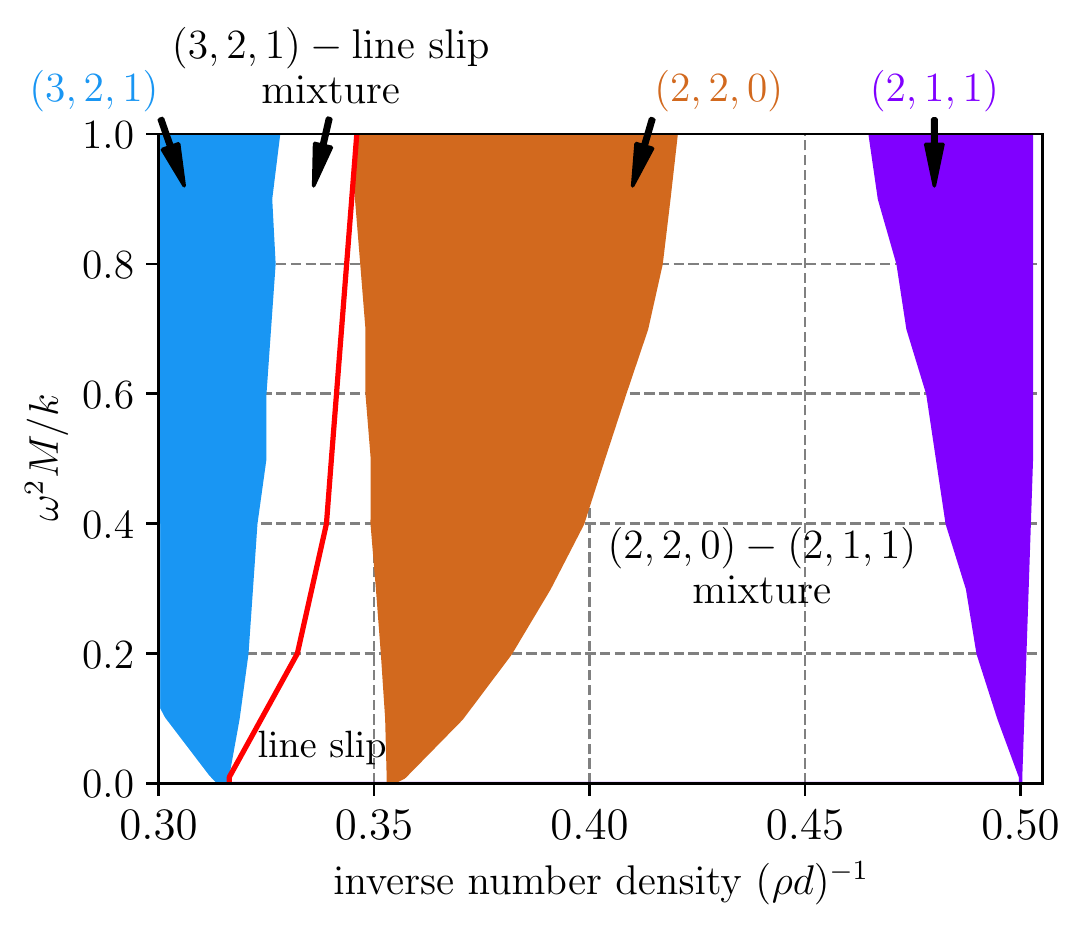}
\caption{
Numerically computed phase diagram from the finite-size simulation with $N = 24$ spheres. 
In addition to the expected homogeneous structures and binary mixture of $(2, 2, 0)$ and $(2, 1, 1)$, a line-slip arrangement, as well as its binary mixture with $(3, 2, 1)$ are found.}
\label{fig:SimPhase}
\end{center}
\end{figure}

An example of our numerical results for a low rotational velocity (i.e., $\omega^2 M/k=0.2$) is shown in Fig \ref{fig:EnergiesSim}.
Here we explore the region between the $(2, 2, 0)$ and $(3, 2, 1)$ homogeneous soft sphere structures.
The blue and brown solid curves are their analytically computed energies; the solid black line is the common tangent between them.
The numerically computed energy, shown by the red dotted line, closely matches the analytic theory within the ranges of the homogeneous phases.
However, a notable difference arises for $0.3205 \le (\rho d)^{-1} \le 0.3520$.
Here, the computed energy is slightly higher than that of the common tangent (binary mixture) due to finite-size effects.
Between the dashed vertical blue [$(\rho d)^{-1} = 0.3205$] and red lines [$(\rho d)^{-1} = 0.3333$] we find a mixture of the $(3, 2, 1)$ uniform structure and a line slip.
Between the dashed vertical red [$(\rho d)^{-1} = 0.3333$] and the brown lines [$(\rho d)^{-1} = 0.3520$] we observe only the line slip.

From our finite-size simulations we compute a limited phase diagram, shown in Fig \ref{fig:SimPhase}, to be compared with Fig \ref{fig:phasediagram}.
In the case of the $(2,1,1)$ and $(2,2,0)$ structures the intervening region is occupied as expected by the $(2,2,0)-(2,1,1)$ mixed phase structure. However, in the case of the $(2,2,0)$ and $(3,2,1)$ structures, the intervening region is split into two parts, featuring the line slip mentioned above for low values of $\omega^2 M / k$.

In part these results corroborate those of the analytic treatment -- in particular as regards the homogeneous phases -- but the intervention of the line slip was an unexpected effect of finite size.
It is to be expected that line slips will play a role in all the other parts of the phase diagram, in finite simulations.
It remains to explore this wide range of possibilities, as well as  the asymptotic trend as $N$ goes to infinity.  
Preliminary results for the case presented here, using $N = 48$ and $96$, indicate that the binary mixture of two homogeneous phases is recovered in that limit.
There also remains the case of hard wall boundary conditions at both ends in a finite sample, which is more directly relevant to the present experiments.

\emph{Conclusion:} The phase diagram presented in Fig \ref{fig:phasediagram} provides an analytic guide to the expected occurrence of equilibrium structures in long rotating columns on the basis of a generic soft sphere model.

We have adduced results from more general simulations.
These introduce interesting features:
They are attributed to finite-size effects, which should be looked for in future experiments.
In any such experiments, one should also be aware of the existence of metastability and hysteresis in macroscopic systems, which we have recently explored in a related context \cite{winkelmann2017simulation, hysteresis2018}.

As noted by Lee \textit{et al.} \cite{lee2017non} this method of creating columnar structures may offer practical applications, particularly if structures can be solidified and continuously extracted, as in the microfluidic experiments of Andrieux \textit{et al.} \cite{wiebke2018}.

\acknowledgments
\emph{Acknowledgments:} This research was supported in part by a research grant from Science Foundation Ireland (SFI) under Grant No 13/IA/1926 and from an Irish Research Council Postgraduate Scholarship (Project ID No. GOIPG/2015/1998). We also acknowledge the support of the MPNS COST Action MP1305 `Flowing matter' and the European Space Agency ESA MAP Metalfoam (AO-99-075) and Soft Matter Dynamics (Contract: 4000115113).
AM acknowledges the TCD Visiting Professorships and Fellowships Benefaction Fund.

We would like to thank the anonymous reviewers for many valid suggestions and comments.

\bibliographystyle{nonspacebib}

\begin{thebibliography}{10}

\bibitem{charbonneau2018}
J.~Norman, E.~Sorrell, Y.~Hu, V.~Siripurapu, J.~Garcia, J.~Bagwell,
  P.~Charbonneau, S.~Lubkin, and M.~Bagnat, Philos. Trans. R. Soc. B
  \textbf{373} (2018).

\bibitem{erickson1973tubular}
R.O. Erickson, Science \textbf{181}, 705 (1973).

\bibitem{brinkley1997microtubules}
W.B. Brinkley, Journal of Structural Biology \textbf{118}, 84 (1997).

\bibitem{amir2012dislocation}
A.~Amir and D.R. Nelson, Proc. Natl. Acad.Sci. U.S.A.
  \textbf{109}, 9833 (2012).

\bibitem{chopra1995boron}
N.~Chopra, R.~Luyken, K.~Cherrey, V.~Crespi, M.~Cohen, S.~Louie, and A.~Zettl,
  Science \textbf{269}, 966 (1995).

\bibitem{smalley2003carbon}
R.E. Smalley, M.S. Dresselhaus, G.~Dresselhaus, and P.~Avouris, \emph{Carbon
  nanotubes: synthesis, structure, properties, and applications}, Vol~80
  (Springer Berlin, 2003).

\bibitem{sanwaria2014helical}
S.~Sanwaria, A.~Horechyy, D.~Wolf, C.Y. Chu, H.L. Chen, P.~Formanek, M.~Stamm,
  R.~Srivastava, and B.~Nandan, Angewandte Chemie International Edition
  \textbf{53}, 9090 (2014).

\bibitem{douglass2017stabilization}
I.~Douglass, H.~Mayger, T.~Hudson, and P.~Harrowell, Soft Matter \textbf{13},
  1344 (2017).

\bibitem{lohr2010helical}
M.A. Lohr, A.M. Alsayed, B.G. Chen, Z.~Zhang, R.D. Kamien, and A.G. Yodh, {Phys
  Rev E} \textbf{81}, 040401 (2010).

\bibitem{meagher2015experimental}
A.~Meagher, F.~Garc{\'\i}a-Moreno, J.~Banhart, A.~Mughal, and S.~Hutzler,
  Colloids Surf., A
  \textbf{473}, 55 (2015).

\bibitem{yamchi2015helical}
M.Z. Yamchi and R.K. Bowles, Physical Review Letters \textbf{115}, 025702
  (2015).

\bibitem{mughal2011phyllotactic}
A.~Mughal, H.K. Chan, and D.~Weaire, Physical Review Letters \textbf{106},
  115704 (2011).

\bibitem{mughal2012dense}
A.~Mughal, H.K.~Chan, D.~Weaire, and S.~Hutzler, Physical Review E \textbf{85},
  051305 (2012).

\bibitem{mughal2013screw}
A.~Mughal, Philosophical Magazine \textbf{93}, 4070 (2013).

\bibitem{mughal2014theory}
A.~Mughal and D.~Weaire, Physical Review E \textbf{89}, 042307 (2014).

\bibitem{fu2016hard}
L.~Fu, W.~Steinhardt, H.~Zhao, J.E. Socolar, and P.~Charbonneau, Soft Matter
  \textbf{12}, 2505 (2016).

\bibitem{fu:2016wm}
L.~Fu, C.~Bian, C.W. Shields~IV, D.F. Cruz, G.P. L{\'o}pez, and P.~Charbonneau,
  Soft Matter \textbf{13}, 3296 (2017).

\bibitem{wood2013self}
D.~Wood, C.~Santangelo, and A.~Dinsmore, Soft Matter \textbf{9}, 10016 (2013).

\bibitem{winkelmann2017simulation}
J.~Winkelmann, B.~Haffner, D.~Weaire, A.~Mughal, and S.~Hutzler, Phys Rev E
  \textbf{97}, 05992 (2017).

\bibitem{hysteresis2018}
A.~Mughal, J.~Winkelmann, D.~Weaire, and S.~Hutzler, Phys Rev E \textbf{98},
  043303 (2018).

\bibitem{lee2017non}
T.~Lee, K.~Gizynski, and B.~Grzybowski, Advanced Materials \textbf{29}, 1704274
  (2017).

\bibitem{basinhopping}
D.J. Wales and J.P.K. Doye, The Journal of Physical Chemistry A \textbf{101},
  5111 (1997).

\bibitem{wiebke2018}
S.~Andrieux, W.~Drenckhan, and C.~Stubenrauch, Langmuir \textbf{34}, 1581
  (2018).

\end{thebibliography}

\end{document}